# Women of the Future in the RAS

**Karen Masters** wonders what the next 100 years will bring for women in astronomy in the UK.

After this year of looking back and celebrating 100 years of women in the RAS, we now ask: what might the future hold? Extrapolating current trends, when might we expect equality in the genders of RAS members, speakers at meetings, award winners and more? Ultimately, when might we stop needing to talk about women in astronomy at all – when it will be as irrelevant to the conversation about astronomy as being a male astronomer is?

I heard all imaginable jokes about being a "woman of the future" when I won the science category of the Women of the Future Awards in 2014, but I think it's a good title to focus on how women stand in the RAS, now and into the near future. At the 2014 Women of the Future Awards ceremony, the most thought-provoking acceptance speech imagined a time when awards celebrating successful young women in the UK were no longer needed; as I write this I share that sentiment. I can't help but wish we weren't still having these same conversations.

**Slow Progress**

It's now roughly 20 years since I started my journey to become a professional astronomer, and the pace of progress towards genuine equity and inclusion in astronomy feels frustratingly slow. If a physics A-level continues to be required for entry into physics/astrophysics degrees we will be stuck with a substantially uneven gender demographic of young professional astrophysicists for quite some time. In 2016, 21% of those obtaining an A-level in physics were female (IoP 2016); this hasn't changed significantly since 1985; and looking back 60 years to 1951, 13% of those studying physics were female (Smithers & Robinson 2006). Fitting a straight line to this 60 year trend and dangerously extrapolating the poor linear fit into the future, we find that we can't expect gender equality in physics A-level until 2163. However this is a generous estimate given that there has actually been zero change in the past 30 years: a trend fit to those data would suggest equality in gender numbers in physics A-level will never be reached.

The stubbornly constant fraction of ~20% of physics A-levels obtained by young women over the last 30 years shows that it's not simply a matter of waiting for women to "work their way through the system" – at least not in the UK. Here the gender make-up of the incoming pool for undergraduate physics and astrophysics courses has not shifted at all, despite significant efforts to engage with young women in both physics and astronomy outreach. We have to hope that work by the Institute of Physics to investigate the root cause of this imbalance (e.g. IoP 2012) will suggest new and innovative ideas to tackle the issue. The new WISE programme "People Like Me" is a positive attempt to use research to inform interventions, suggesting a focus on the types of people who study physics, rather than just showing what physicists do.

In the US, where there's no subject narrowing until university level, approximate gender parity in undergraduate astronomy degrees has been reached, with a roughly constant proportion of ~40% women from 2003 to 2012 (Mulvey & Nicholson 2014), although physics degrees remain at just 20% women. This suggests that radical ideas like abolishing the requirement for A-level physics for physics and engineering degrees (Coles 2015), could cause a swift cultural change.

Let's be clear that the gender imbalance in physics and astrophysics is a cultural or societal effect, as figures on the representation of women in astronomy internationally (from the International Astronomical Union), clearly show. In 2015, just 13% of the UK's IAU members were women, while 23% of Turkish members, 25% of Italian, and 39% of Argentine members are female (IAU 2015).

**Inconsistent imbalance**

The gender imbalance is also field-specific. There is no gender imbalance at the undergraduate level today in many areas of science, and there is evidence that cultural beliefs about the innate talent needed to succeed in certain fields may be to blame (Leslie *et al.* 2015). Campaigns like "Let Toys Be Toys" and "Pink Stinks" remind us that, in the world of children in 2016, gender is still often considered the primary distinguishing characteristic. As a mother of primary- school-age children I'm astonished at how often things like "that's just how boys play" or "of course, that's because she's a girl" are said.

Caring responsibilities remain stubbornly gendered with mostly female primary-school teachers, and mostly mothers on the school run and in school support groups such as Parent Teacher Associations. This disadvantages both mothers in astronomy, who may be pressured to maintain unrealistic work–life balances, but also the fathers, who may feel less able to take the paternity leave or flexible working patterns they are entitled to.

We must also recognize that the infamous leaky pipeline is still at work. The most recent RAS demographics surveys still show a declining fraction of women with career seniority over and above what would be expected just from normal progression. The 1980s and 1990s were decades in which women and girls were encouraged to "have it all"; in reality, structural problems with careers, such as young scientists being encouraged to move frequently all over the world, or to be hugely productive in their peak child-rearing years, disproportionately affect both the ability and the desire of young women to remain in academia (Ivie *et al.* 2016).

Even the classic two-body problem also disproportionately affects women. Run the numbers in your head and you can see that, in a field with four times as many men as women, a higher proportion of women there are going to be married to male astronomers than men married to female astronomers.

Female astronomers are undeniably under-represented in leadership roles; this has recently been shown in my own collaboration, SDSS-IV, which is now working hard to make positive changes (Lundgren *et al.* 2014). There are other problems in astrophysics today that look likely to propagate this structural issue many years into the future, such as female astronomers finding it harder to win observing time, demonstrated by Reid (2014) for the

Hubble Space Telescope, and Patat (2016) for ESO facilities, obtaining on average 10% fewer citations for their papers (Caplar, Tacchella & Birrer 2016), or feeling less able to ask questions at research conferences (Pritchard et al. 2014).

**Sexual Harassment**

Frankly, 2016 has been an uncomfortable year to be a woman in astronomy, and hasn't always left me feeling hopeful for the future. You must have been living under a rock if you've missed the series of high- profile sexual harassment cases coming out in the US astronomical community (e.g. Witze 2016). We should all recognize a debt of gratitude for the young women who risked their reputations and careers to come forward to report such behaviour. Their bravery will make our field a safer and more welcoming one for all young people to enter in future. However, I am disheartened that there are still astronomers who seem less concerned about the impact on these young astronomers (and the many more who will have chosen not to come forward), and more concerned to make sure the impact of the behaviour isn't too great on the perpetrators' careers, or the reputation of our field.

As a community we should take care that we do not worry more about protecting senior (mostly) men against false accusation than supporting innocent victims. False accusations have been repeatedly shown to be rare (e.g. Lisak *et al.* 2010) and yet most processes seem to be set up to protect against them more than to support victims. If you take one thing away from this article it's this: if a young astronomer comes to you with a story of harassment, you should not in that moment express doubt, nor ask for any proof, but instead trust the process to find the truth and offer your support and sympathy. A simple "I'm sorry that has happened to you" can mean the world. You need not worry that people making false accusations will get away with it – the processes will ensure they will not – but rather make sure you do not miss that unique opportunity to express sympathy and support to a genuine victim.

Our field can show its strength, humanity and inclusive practices by supporting the victims and outing the perpetrators. We need to make it very clear this behaviour is not acceptable. I don't think I've been in a group of astronomers this year where these cases of sexual harassment in astronomy haven't come up at some point. There have been no high-profile cases in the UK in which a complaint has been upheld, but we should be careful not to be too smug. We can hope this means no-one in the UK community has been harassing young astronomers, but we should worry that all it means is that they have yet to be publicly exposed.

People are people in all careers and countries, and while most will act professionally, bad behaviour can happen anywhere there are people. A disturbing article for the *Guardian* (Weale & Batty 2016) suggested that non-disclosure clauses in the UK HE system are common following sexual harassment cases, so officially reported harassment is routinely hidden – hardly a reassuring conclusion.

On the positive side, such behaviour is no longer considered acceptable and unavoidable. Initiatives such as Athena SWAN and Project Juno provide a framework for physics and astronomy departments to critically assess their environment and culture and offer suggestions for how to counteract not just obvious bad behaviour, but also ingrained and systemic biases, such as unconscious bias. They also make suggestions to support those with parenting duties, such as making sure meetings are not held out of core hours – the timing of the RAS Ordinary Meetings from 4 to 6 p.m. on Friday is not a shining example here!

Social media has changed the way information is spread, in good ways as well as bad. Social media is just a communication tool and its strengths are also its weaknesses: the ease with which minority and/or excluded communities can find each other through hashtags and keywords also makes it easy for those who would wish to silence them to find them. Those who engage with social media are a community like any other and have to work as hard to create the right culture and environment as the offline society.

**Conclusion**

As you can read elsewhere in this issue, the RAS membership included just 1.6% women in 1916. This has increased to 16.7% women today, up 15% points in 100 years. Let's hope we can manage better in the next 100 years: I'd like to see more than 20% of women taking physics at A-level. And I'd like to see far more than 20% of the membership identifying as women in 2116. All the signs are there that we can work together to achieve that.

---

**Author**
Karen Masters is a Reader in Astronomy and Astrophysics at the Institute of Cosmology and Gravitation, University of Portsmouth.